\documentclass[aip,jcp,reprint]{revtex4-1}

\usepackage{gensymb}
\usepackage{amssymb}
\usepackage{color,graphicx}
\usepackage{subfig}
\usepackage{transparent}
\usepackage{textcomp}
\usepackage{siunitx}
\usepackage{braket}

\begin{document}

\setcounter{secnumdepth}{3} 


\title{Handshake electron transfer from hydrogen Rydberg atoms incident at a series of metallic thin films.}


\author{J. A.  Gibbard}
\affiliation{Department of Chemistry, University of Oxford, Chemistry Research Laboratory, Oxford OX1 3TA, United Kingdom}
\author{T. P. Softley}
\affiliation{University of Birmingham, Edgbaston, Birmingham, B15 2TT, United Kingdom}


\date{\today}

\begin{abstract}
Thin metallic films have a 1D quantum well along the surface normal direction, which yields particle-in-a-box style electronic quantum states.  However the quantum well is not infinitely deep and the wavefunctions of these states penetrate outside the surface where the electron is bound by its own image-charge attraction. Therefore a series of discrete, vacant states reach out from the thin film into the vacuum increasing the probability of electron transfer from an external atom or molecule to the thin film, especially for the resonant case where the quantum well energy matches that of the Rydberg atom.
 We show that `handshake' electron transfer from a highly excited Rydberg atom to these thin-film states is experimentally measurable. Thicker  films, have a wider 1D box, changing the energetic distribution and image-state contribution to the thin film wavefunctions, resulting in more resonances.  Calculations successfully predict the number of resonances and the nature of the thin-film wavefunctions for a given film thickness.
%

\end{abstract}
\maketitle

\section{Introduction}

The particle in a 1D box problem is a classic exercise in elementary quantum mechanics, but is often taught as a hypothetical situation that presents a tractable solution of the Schr\"{o}dinger equation.  Nevertheless it is useful for calculating translational partition functions in statistical mechanics (and hence thermodynamic properties of gases) and is also applicable as a zero-order model for various other real physical problems. One of these is the case of a thin metallic film, which has energy states and wavefunctions that to a first approximation correspond to those of a particle in a 1D  box. In this paper we present an experiment in which an external Rydberg H atom - with its electron excited into a  high-energy loosely bound state -  impinges on a thin metallic film and transfers an electron into the 1D box.  
The transfer is aided by the protrusion of the vacant thin-film states into the vacuum such that at long range the Rydberg electron wavefunction of the incoming atom overlaps with the thin-film wavefunction and the electron is handed over to be captured into the 1D box - hence we refer to this as `handshake electron transfer'.

Usman \textit{et al} used a quantum wavepacket propagation approach to theoretically investigate  electron transfer of this type from a 
ground state H$^{-}$ ion 
to a metallic thin-film \cite{thin_films}. Their calculations predicted that when the electronic energy of the H$^-$ was close to that of a 1D thin-film state reaching into the vacuum, the rate of electron transfer would be enhanced by a resonance effect.
However the advantage of using Rydberg atoms  over H$^-$ ions is that a larger energy range of the thin-film states can be probed, due to the experimental accessibility of Rydberg states with different principal quantum numbers over a wide range of energies. In addition, there is better matching of the spatial extent of the Rydberg wavefunction with the protruding wavefunction from the thin-film state. 

Many previous studies have investigated electron transfer from a Rydberg atom to a flat metallic surface where the projected electronic bands of the surface are degenerate with the experimentally accessible quantum states of the Rydberg atoms. In this case the characteristics of the electron transfer process are dominated by the properties of the Rydberg state (principal quantum number $n$ and parabolic quantum number $k$) and the experimental conditions (applied field and Rydberg atom collisional velocity) \cite{eric_prl, wpp_atoms,stray_dunn,dunn3}. 
However other studies using projected band gap surfaces \cite{Cu_theory, Cuexpt}, doped silicon surfaces \cite{Silicon}, insulating thin films \cite{mccown}, dielectric materials \cite{dielectric} and adlayers indicate that the surface ionization of a Rydberg atom can also be used  to learn about the geometrical or electronic structure of the surface.

In a zero-order approximation, treating the thin film as a 1D  well of infinite depth, the electronic states are fully localized within the thin film, have a nodal structure described by a quantum number $n_{\mathrm{well}}$ and become increasingly widely spaced with increasing energy. 
As the film thickness increases, so does the width of the well resulting in lower-energy well states which are closer together. 
Parallel to the surface, the thin film behaves as a 2D metallic continuum. Each discrete well state forms the lower bound for a series of 2D bands in the metallic thin film,  such that as the film tends towards infinite thickness, the bulk limit of a 3D band is recovered. 
Previous photoemission studies of thin films have focused on determining the energies and spatial distribution \cite{cuthinfilm} of the well states within thin films, and their dependence on the substrate and the evaporant \cite{quantwellrev, quantwellrev2}.

\begin{figure*}[t]
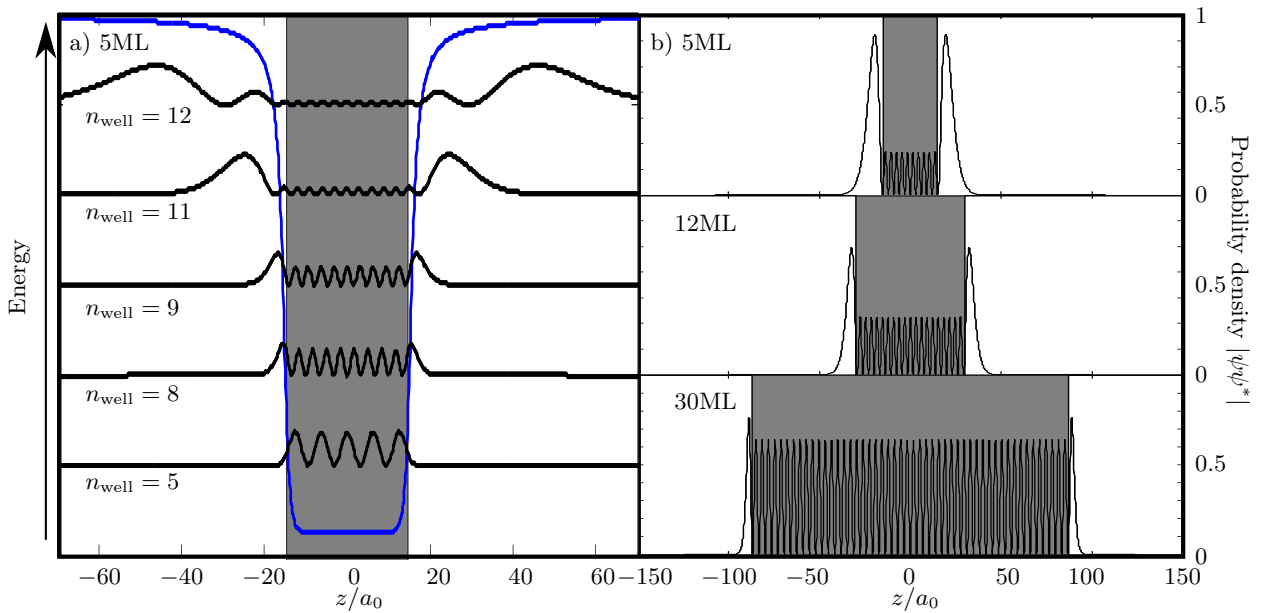
\caption{a) A selection of 1D wavefunction probability densities for the five-monolayer thin films ordered according to increasing energy (artificially spaced on the energy axis for clarity). The blue line depicts the 1D quantum well. b) The probability density of a  five-, twelve- and thirty-monolayer thin-film wavefunction, near the unperturbed energy of the first image state. The gray box indicates the physical thickness of the film and the $z$ coordinate is the distance from the center of the thin film.  }
\label{fig:tf_wfs}
\end{figure*}

However, the 1D quantum well is not of infinite depth, hence penetration of the wavefunction outside the well can occur into the image-charge region. Classically, an electron near a metallic surface induces image charges within the 
surface.  
The Coloumb-like attraction between the  electron and its image charge yields a discrete Rydberg-like series of bound states protruding from the surface into the vacuum. For the surface of a bulk material, the energies of these image states are given by (in a.u.)
\begin{equation}
E_{\mathrm{IS}}(n_{\mathrm{img}})=-\frac{1}{16}\frac{1}{2(n_{\mathrm{img}}+a)^{2}},
\label{eq:imens}
\end{equation}
where $n_{\mathrm{img}}$ is the image-state quantum number and $a$ is the quantum defect parameter for a given surface. 
The pure image states extend far into the vacuum and so significant overlap is possible with the electronic wavefunction of an incoming Rydberg atom at long range. 
When image states are degenerate with the surface conduction band of a bulk metal surface they are mixed with the conduction band and broadened, such that an individual image state is no longer measurable, but instead all conduction-band states
have some image-state character, and are known as image resonances. 
For a thin film, the conduction band is not fully continuous and therefore the image states are admixed into those discrete 1D well states that have nearby energies.
Thus, the thin-film states,  which during handshake electron transfer can accept the Rydberg electron, have combined character of the image states and the 1D-well states.  Figure \ref{fig:tf_wfs} shows the probability density for a selection of these thin-film states, referred to in this paper as `well-image' states,  which have an image-state tail protruding into the vacuum on both sides of the thin film, and particle-in-a-box style nodes within the film itself. 
The energy levels and wavefunctions of a single electron for the thin film system were determined in this work via diagonalization of the Hamiltonian for a one-dimensional Jellium quantum well, shown in figure \ref{fig:tf_wfs}a, on a Sinc DVR  grid \cite{sinc1}. 
\begin{figure*}
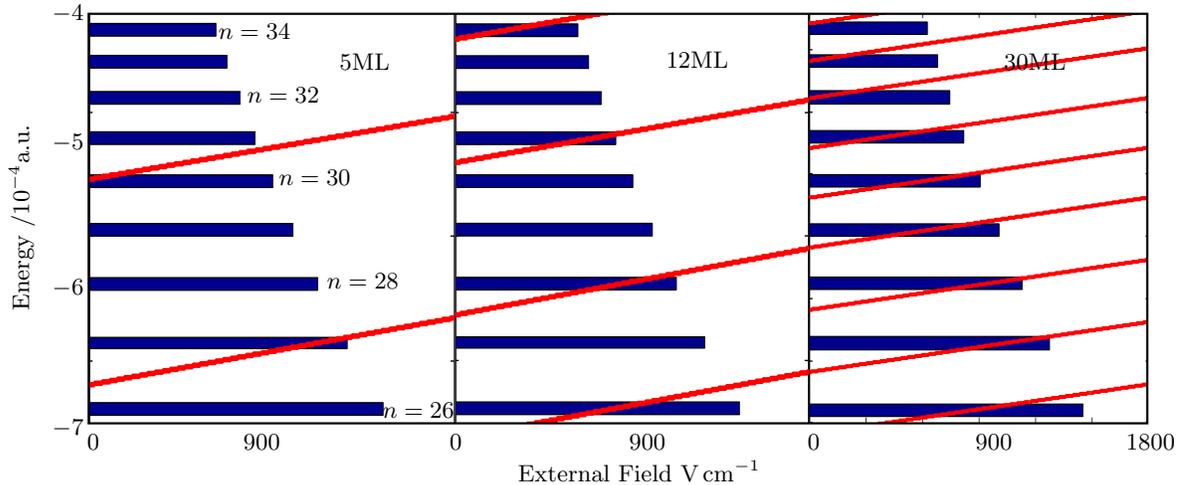
\caption{The calculated energy levels of the (a) five-monolayer, (b) twelve-monolayer and (c) thirty-monolayer thin-film 1D quantum well are shown in red, and the $n=26-34$ $k=0$ H atom Rydberg states. The thickness of the blue line indicates the shift in Rydberg energy on approach to the surface, and the length of the line indicates the range of applied fields over which surface ionization can be measured before the onset of field ionization. 
}
\label{fig:tfens}
\end{figure*}

The states shown in figure \ref{fig:tf_wfs} are calculated for a five-monolayer (5ML) thin film.   The lowest energy wavefunctions, at the 
bottom of the well (e.g. the $n_{well}=5$ state in Fig 1), have almost pure particle-in-a-box character, with a clearly defined number of nodes and little protrusion into the vacuum. 
At these lowest energies there is negligible contribution from the image states, as the zero-order  image-state series is very widely spaced.
As the energy increases the thin-film states develop increasing image-state character, with significant probability density outside of the surface. At the experimentally accessible energy range of the H atom Rydberg states, the wavefunctions have several nodes outside the surface and protrude for nanometers into the vacuum;
 however there is still well-defined particle-in-a-box character and the size of the contribution from well or image character varies from one wavefunction to the next. Near the vacuum level, the  states have almost pure image-state character, with little amplitude in the well region (e.g., $n_{well}= 11$ or 12.)

As the film thickness increases, the number of zero-order well states increases due to the larger well width, but the number of zero-order image states remains the same, resulting in thin-film wavefunctions which have proportionately more well-state character. For thicker films there are more pure low-energy well states than for the five-monolayer case, and less pure-image states near the vacuum level.
Figure \ref{fig:tf_wfs}b plots the probability density of the well-image state nearest to the zero-order energy of the first image state, for a five-, twelve- and thirty-monolayer film. As the film thickness increases, the amplitude of the wavefunction outside the film decreases, as the state has more well character and the electron is more likely to be found within the physical boundary of the film. Ultimately as the film thickness is increased towards infinity, the bulk 3D limit of a band of image resonances is reached. 
  
Handshake electron transfer, where we consider the thin-film state as extending an empty hand into the vacuum to collect a degenerate incoming Rydberg electron, is more efficient when the film states  can reach further into the vacuum. Therefore for thinner films, where the states have greater image-state character, electron transfer is expected to occur more readily.

Figure \ref{fig:tfens} shows the calculated energy levels of a five-, twelve- and thirty-monolayer thin film in red, and the $n=26-34$ $k=0$ H atom Rydberg states in blue.
The diagonalizations were cycled over the range of experimental applied fields used in this work, as states with image character are susceptible to external fields. 
Where the red and blue lines intersect, resonance-enhanced electron transfer is expected.
Thicker films have a denser manifold of well-image states due to the larger zero-order well state contribution, and so more resonances are expected. 

Once the energetic spacings between thin film and Rydberg states become comparable, the system goes below the effective resolution of our experiment, and the thin film becomes indistinguishable from a `bulk' surface. This loss of resolution is expected to occur at around thirty monolayers, or 4 $\si{\nano\meter}$ thickness, as shown in figure \ref{fig:tfens}c. It is expected that more resonances will be seen in the twelve-monolayer case than the five-monolayer case, and resonant energies will shift between the two thin films. 

\section{Experimental Results}

\subsection{Methods}

\begin{figure}
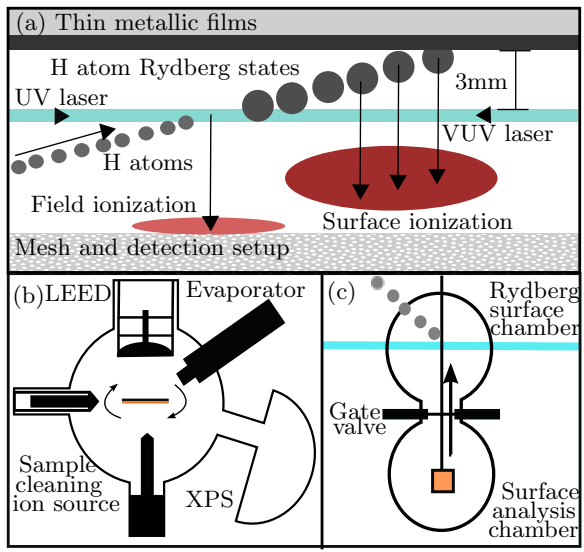
\caption{a) Two-color excitation of H atoms produces a beam of Rydberg atoms to probe a thin metallic-film surface. Electron transfer to the surface yields ions which are measured as a function of extraction field. b) Schematic of the surface analysis chamber. c) The xyz manipulator allows the surface to be moved between the surface analysis chamber and the Rydberg-surface experiment under vacuum.}
\label{fig:exp_setup}
\end{figure}

The experimental setup is shown in figure \ref{fig:exp_setup} and has been described in more detail elsewhere  \cite{Cuexpt, eric_prl}.  In outline, a velocity- and quantum-state-selected beam of Rydberg H atoms is incident on the surface of an iron thin film deposited in vacuum on a Muscovite substrate.  The ions (protons) produced as a result of electron transfer to the surface are extracted away from the surface to a detector by an applied field of variable magnitude.  
Surface-ionization profiles, depicting the number of ions produced via surface ionization  as a function of applied field, were measured for H atom Rydberg states $n=26-34$ for three deposited iron films of different thicknesses. As the field is increased, ions that have been formed closer to the surface can be extracted, and therefore the profile reflects the range of distances over which ionization occurs \cite{Cuexpt}.

The metallic films are prepared {\it in situ} in the lower chamber shown in Fig. 3, for optimal control of the flatness and cleanliness. An air-cleaved muscovite crystal is used as the evaporation substrate, and its flatness is confirmed by LEED before deposition. An electron-beam evaporator with an integrated flux monitor is used to deposit iron onto the substrate. The thickness of the film is determined from the flux, and confirmed via XPS measurements and subsequent intensity analysis. 
After the thin films have been prepared in the surface-analysis chamber the surface is moved under vacuum into the Rydberg-surface collision chamber.  

To produce the H atoms, a beam of pure ammonia is created by a pulsed nozzle ($10\si{\hertz}$), and passes through a capillary mounted on the nozzle where the ammonia is photodissociated using an ArF excimer laser at $193\si{\nano\meter}$, to produce the hydrogen atoms. After traveling a distance of {\it ca} $50\si{\centi\meter}$, the hydrogen atoms are excited to the $2p$ state via a VUV photon at the Lyman-$\alpha$ wavelength $121.6\si{\nano\meter}$, which is formed by passing a frequency-doubled (5 ns pulse length)  dye laser beam through a rare-gas  (Ar/Kr) tripling cell.  A counter-propagating tunable frequency-doubled  dye laser  excites the atoms to the selected $n=26-34$ Rydberg states  with a UV photon in the range $365.8 - 366.7\si{\nano\meter}$. The excitation occurs in the presence of a small applied electric field to allow spectroscopic separation of states with different polarization with respect to the surface \cite{eric_prl}. When the excitation to the Rydberg state occurs in the presence of a small applied field, Stark states are formed. A Stark state is characterized by its parabolic quantum number $k$, and polarization of the electron density along the field axis. The effect of the surface potential on the electron density is similar and means that,
 with the field direction pointing away from the surface, red-shifted Stark states are more readily surface ionized than their blue-shifted counterparts \cite{eric_prl}.  
  In this work the $k=0$ state is used throughout as its relatively uniform electron density makes it less sensitive to applied electric fields.  The lasers and molecular/atomic beam are pulsed and synchronized at $10 \si{\hertz}$.

After surface-ionization has occurred, the resulting protons are extracted by a large applied field (100 - 2000 Vcm$^{-1}$); without this field the ions would fly towards the surface to be neutralized via an Auger process \cite{Cazalilla:1998p36} and the surface-ionization signal would be unmeasurable. 
The minimum field required to extract an ion $F_{\mathrm{min}}$ depends on the surface-atom separation $D$ at which surface ionization occurs and the Rydberg atom's kinetic energy  along the surface normal direction $T_{\perp}$,
\begin{equation}
F_{\mathrm{min}}(D, T_{\perp})=\Bigg[ \frac{1}{2D}+\sqrt{\frac{T_{\perp}}{D}}\Bigg]^{2}
\label{eq:fmin}
\end{equation}
As the applied field is increased, ions produced via surface ionization closer to the surface are extracted. The ions are focused by a series of ion optics and detected by a microchannel plate, coupled to a phosphor screen and monitored by a photomultiplier tube. The ion signal is measured on an oscilloscope as a time-of-flight trace so that the integral of the surface-ionization signal peak is recorded for every applied field. 

Typical `surface-ionization profiles' where the surface-ionization signal is recorded as function of applied field (and thus indirectly, ionization distance from the surface) are shown in figure \ref{fig:5tf_sps}.  At sufficiently large fields the Rydberg atoms are field-ionized immediately on application of the extraction field and before interaction with the surface. These ions appear at a different time of flight, hence are readily distinguished from those produced by surface ionization. The decreasing signal at high field  in the surface-ionization profile is thus attributable to the competition with field ionization. The field-ionization signal, observed  earlier in time, can be used to normalize the surface-ionization signal to account for fluctuations in laser power and atomic beam density.

The results are recorded for a range of different collision velocities perpendicular to the surface - $500\si{\meter\per\second}$, $600\si{\meter\per\second}$, $700\si{\meter\per\second}$ and $850\si{\meter\per\second}$.  The spreading of the atomic beam over the 50 cm path length between the photolysis and the Rydberg excitation, is such that only a narrow velocity component (approximately 1$\%$ velocity resolution) is selected in the excitation - the mean velocity can therefore be adjusted by varying the time delay between photolysis and excitation.  

Some low-intensity surface-ionization signal, measurable only near the field-ionization limit, was recorded for H atoms incident at a bare muscovite surface. 
The surface-ionization profiles for bare muscovite
are very different from those when thin films are present and it is clear that once the iron film has been deposited, the surface-ionization behaviour is dominated by the interaction between the Rydberg atom and the metallic film only.

\subsection{Results}
\begin{figure}
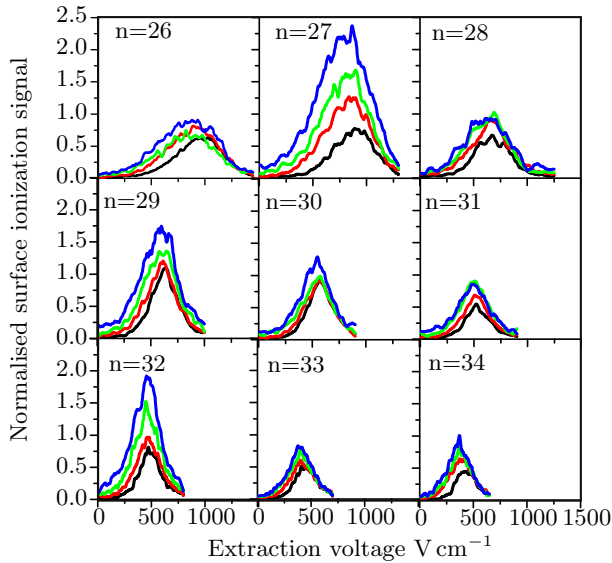
\caption{The surface-ionization profiles for the $k=0$ $n=26-34$ hydrogen atom Rydberg states incident at a five-monolayer iron thin film for a series of perpendicular collisional velocities: $850\si{\meter\per\second}$ (black); 700 ms$^{-1}$ (red); 600 ms$^{-1}$ (green); and 500 ms$^{-1}$ (blue).}
\label{fig:5tf_sps}
\end{figure}

Figure \ref{fig:5tf_sps} plots the surface-ionization profiles for $n=26-34$ hydrogen Rydberg atoms incident at a five-monolayer iron film evaporated onto muscovite.
The surface-ionization profiles for $n=27,~29$ and  $32$ show  a significantly higher surface-ionization signal than the others, indicating surface ionization is resonantly enhanced and occurring when the atom is further from the surface in these cases. 
The resonance-enhancement of the charge transfer occurs when the Rydberg energy matches the zero-parallel-momentum ($k_{\parallel}=0$) thin-film   states at the base of each energy band in the full 3D problem - the enhancement reflects a marked propensity for the electron flux to transfer perpendicularly to the surface  so as to conserve angular momentum. 
After transfer, the population subsequently disperses amongst the degenerate higher parallel momentum states associated with lower energy 1D-well states.
The energy-level diagram in figure \ref{fig:tfens}a predicts resonances at $n=27,~30,~35$ for a five-monolayer thin film, which is a good qualitative match to the experimental results, accurately predicting the number of resonances in this region. 

It is also observed that as the perpendicular collisional velocity is decreased, the amount of surface-ionization signal increases. Decreasing velocity results in an increased mean atom-surface separation for surface ionization (the atoms have more time to ionize in a given incremental distance range as they approach the surface), and a greater detectability of the ions (see equation \ref{eq:fmin}) \cite{Cu_theory}. 
For on-resonant states, the velocity effects are exaggerated more than for off-resonant states. In particular, the $n=27$ signal shows a larger variation of integrated surface signal with velocity than its off-resonant neighbouring $n$ values.  
\begin{figure*}
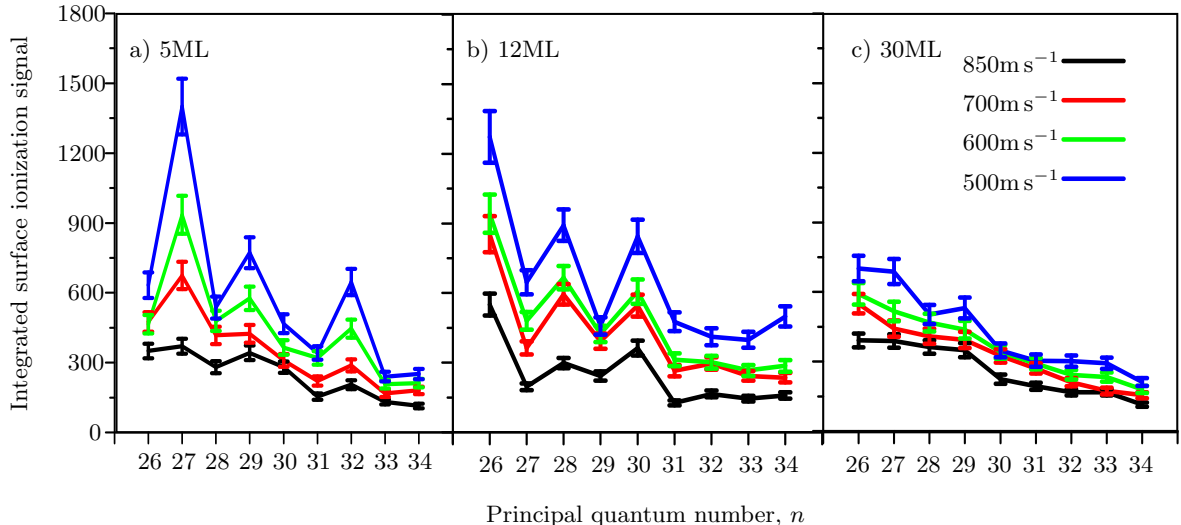
\vspace{-0.5cm}
\caption{The integrated surface-ionization signal of the  $n=26-34$,  $k=0$ hydrogen atom Rydberg states  with a range of perpendicular collisional velocities incident at a five, twelve and thirty-monolayer iron thin film.}
\label{fig:allint}
\end{figure*}

Figure \ref{fig:allint} plots the total detected surface-ionization signal, integrated over the field range, for the $n=26-34$, $k=0$ H atom Rydberg states incident at a series of thin metallic films.  The integrated signal data in the figure show the resonances more clearly, especially at the lowest velocity (blue line).
Thicker films were prepared in vacuo by successive deposition cycles using the evaporator shown in figure \ref{fig:exp_setup}. 
Figure \ref{fig:allint}b shows the integrated surface-ionization signal for the $n=26-34$, $k=0$ hydrogen Rydberg atoms interacting with a twelve-monolayer ($1.6$ nm) film at a range of collisional velocities. 
Resonances now occur at $n=26,~ 28,~ 30$ and $34$ as the thicker film, with its wider well, has different energy levels to the five-monolayer film. 
The observation of the fourth resonance supports the idea of the 1D particle-in-a-box behaviour of the energy levels within the thin film, with the film width determining the separation between the well states (For a 1D box of infinite height walls, $E\propto n^2/L^2$ where L is the size of the box). 
 The calculated energy level diagram shown in figure \ref{fig:tfens}b predicts resonances at $n=26,~28,~31$ and $34$ which is a good match to the experimental results, especially considering the approximations in the theoretical calculations.
The velocity dependence of these surface-ionization profiles mirrors the behaviour for the five-monolayer films with a greater spread of intensity with velocity for the on-resonance principal quantum numbers.

The integrated surface-ionization signal for a thirty-monolayer thin film, shown in figure \ref{fig:allint}c,  depicts a more monotonic variation with $n$ than the other thin films, indicating surface-ionization behaviour more like that of a Rydberg atom at a bulk conducting metal surface. 
This behaviour was predicted by the calculated energy level diagram for the thirty-monolayer system, shown in figure \ref{fig:tfens}c, where the spacings between the thin-film 1D states states are comparable in size to the spacings between the Rydberg-atom energies. As each Rydberg state is now in resonance with a thin-film state, the thirty-monolayer thin film can be thought of as `bulk-like' with respect to the resolution of our experiment, despite the energy levels along the surface normal still not forming a true continuum. 

Figure \ref{fig:allint}c shows that the integrated surface-ionization signal decreases as the principal quantum number is increased, much as it would for a bulk metallic surface such as gold \cite{eric_prl}. Lower-$n$ H atoms field ionize at larger fields, and so the surface-ionization signal can be detected over a greater range of fields than for high-$n$ Rydberg atoms. Therefore the surface-ionization signal is integrated over a larger range of fields, and increases with decreasing $n$. The small variation in integrated signal from one $n$ value to another is likely to be due to either marginally better energy matching between Rydberg and thin-film levels for one $n$ than another, or to variations in the image-state character of the wavefunction; but fundamentally, the energies of all the principal quantum numbers coincide with one or more thin-film states in the field range studied. 
Finally, decreased perpendicular collisional velocity again increases the amount of surface-ionization signal, but by a smaller amount than seen for the other thin-film systems. 

\section{Discussion}
\begin{figure}[t]
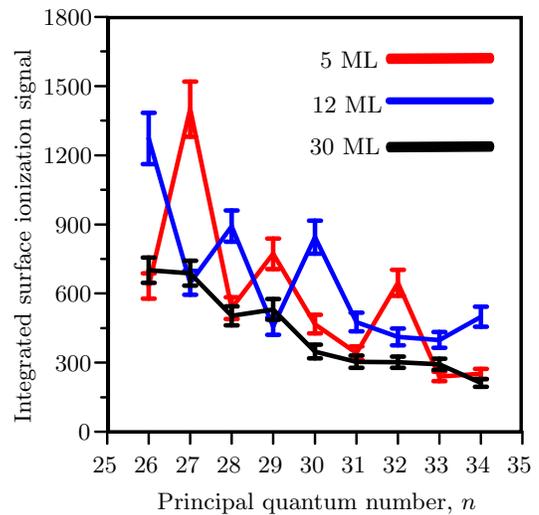
\caption{Integrated  surface-ionization signals from  the ionization profiles of various thin-film thicknesses at a collisional velocity of 500 \si{\meter\per\second}:  five-monolayer film (red); twelve-monolayer (blue); and thirty-monolayer (black).}
\label{fig:tfs_vel_comp}
\end{figure}

Figure \ref{fig:tfs_vel_comp} superimposes the integrated surface-ionization signal for the three thin films at a collisional velocity of $500\si{\meter\per\second}$.  
The blue and red lines, which depict the integrated surface-ionization signal for the five- and twelve-monolayer thin films, have clear peaks at different principal quantum numbers, showing that the thickness of the film affects the energy and distribution of the states reaching into the vacuum to undergo handshake transfer with the Rydberg atom. 
The integrated surface-ionization signal of the thirty-monolayer thin film, shown by the black line, has no distinct peaks and behaves more like a bulk metallic surface, such as gold. A thirty-monolayer film presents the effective  limit of resolution for these experiments, where the density of Rydberg states is comparable to the density of thin-film states. 
The intensity of the integrated  surface-ionization signal obtained for the thirty-monolayer thin film is comparable with signal for the off-resonance Rydberg states for the five- and twelve-monolayer films.
Calculations show that the wavefunctions of the thirty-monolayer thin film have less image-state character, due to the greater contribution of the well-localized states for thicker films.  Effectively these thin-film states do not extend as far into the vacuum, so there is poorer overlap with the incoming Rydberg state wavefunctions and overall less efficient handshake electron transfer.

This work demonstrates that we can take advantage of the wide range of accessible Rydberg energies to probe the surface electronic structure of thin metallic films over a wide energy range. The increases in signal observed at certain principal quantum numbers compared to the surface of bulk metals show how the image-state part of the surface wavefunction receives the electron from the Rydberg atom and into the thin film. This novel approach to characterizing the states of thin films not only illustrates some classic physics but could in principle be refined to yield information  about surface states with the same spectroscopic precision as the Rydberg energy levels can be measured.


\end{document}